\documentclass[pre,twocolumn,groupedaddress,showpacs,showkeys,amsmath,amssymb,floatfix]{revtex4}

\usepackage{graphicx}
\usepackage{dcolumn}
\usepackage{bm}

\begin{document}

\title{Directed negative-weight percolation}
\author{C. Norrenbrock}
\email{christoph.norrenbrock@uni-oldenburg.de}
\author{M. M. Mkrtchian}
\email{mitchell.mger.mkrtchian@uni-oldenburg.de}
\author{A. K. Hartmann}
\email{alexander.hartmann@uni-oldenburg.de}
\affiliation{Institut f\"ur Physik, Universit\"at Oldenburg, 26111 Oldenburg, Germany}
\date{\today}

\begin{abstract}
We consider a directed variant of the negative-weight percolation model in a two-dimensional, 
periodic, square lattice. The problem exhibits edge weights which are taken from a distribution 
that allows for both positive and negative values. Additionally, in this model variant all edges 
are directed. For a given realization of the disorder, a minimally weighted loop/path configuration 
is determined by performing a non-trivial transformation of the original lattice into a minimum 
weight perfect matching problem. For this problem, fast polynomial-time algorithms are available,
thus we could study large systems with high accuracy. Depending on the fraction of negatively and 
positively weighted edges in the lattice, a continuous phase transition can be identified, whose 
characterizing critical exponents we have estimated by a finite-size scaling analyses of the 
numerically obtained data. We observe a strong change of the universality class with respect to 
standard directed percolation, as well as with respect to undirected negative-weight percolation.  
Furthermore, the relation to directed polymers in random media is illustrated.
\end{abstract}

\maketitle

\section{Introduction \label{sect:introduction}}
In statistical physics, one of the central targets is to study systems that exhibit continuous 
phase transitions. Due to the diverging correlation length in the critical region, long-range 
correlations are not affected by details of microscopic interactions, but depend on symmetry 
properties of the underlying model only. For that reason models that exhibit continuous phase 
transition can be grouped in universality classes, which are characterized by a set of critical 
exponents and their functional relations, i.e.\ scaling laws \cite{stanley1999}. One of the most 
basic and intensively studied universality classes is that of standard percolation 
\cite{stauffer1979,stauffer1992}, which addresses the question of connectivity. Based on a tunable 
parameter $p$, sites or links in a given lattice get occupied or stay empty. Then, the central 
objects of interest are clusters consisting of adjacent and occupied sites. Above a certain value 
of $p=p_c$, i.e. the critical point, a lattice-spanning cluster emerges in the thermodynamic 
limit. Even the model is probably the simplest possible model exhibiting a phase transition, its 
importance comes from the following facts. First, it allows to study basically all fundamental 
aspects of phase transitions within a very basic framework. Second, many much more complex phase 
transitions can be traced back to an underlying percolation transition, e.g., the percolation of 
Fortuin-Kasteleyn clusters \cite{fortuin1972} in the Ising model.

In standard percolation, there is no directional information in the connectivity pattern. Thus, not 
surprisingly, the critical exponents describing this phase transition differ from those that 
describe the phase transition in directed percolation (DP) \cite{hinrichsen2000}, which is a variant 
of standard percolation, where the links carry a direction, leading to an anisotropic behavior. Note 
that this directionality can be interpreted as time direction, making DP relevant for the 
description of non-equilibrium processes. In particular, because of the anisotropic nature of the 
cluster building process in DP, correlations are not governed by one but two correlation lengths: 
$\xi_\parallel$ and $\xi_\perp$.

Also, in standard percolation, the links do not carry any weights, thus, one can assume all weights 
being one, i.e., they are in particular  positive. Recently, a percolation model called 
``negative-weight percolation'' (NWP) was introduced \cite{melchert2008}, where random weights are
attached to the links, and, in particular, weights of either sign are allowed. Algorithmically, this 
means special global optimization polynomial-time ``matching'' algorithms have to applied, see Sec.\ 
\ref{sect:modelalgo}. This leads, interestingly, to a new type of behavior giving rise to a 
different universality class compared to standard percolation. In a series of papers
\cite{apolo2009,melchert2010,melchert2011,mezard2011,claussen2012,norrenbrock2013,melchert2014}, NWP
has been studied in different dimensions and different variants.

It has been shown that two distinct phases can be identified depending on a disorder parameter 
$\rho$, which controls the amount of negative weights. (i) for small $\rho$ the geometric objects 
are rather small and straight-lined, which reflects a self-affine scaling, (ii) for large $\rho$ the 
geometric objects scale self-similar and can wind around the lattice. In Ref.\ \cite{melchert2008} 
the disorder-driven phase transition was investigated by means of finite-size scaling analyses and 
it turned out that the critical exponents were universal in 2D (different lattice geometries and 
disorder distribution were studied). Further studies regarding isotropic NWP address the influence 
of dilution on the critical properties \cite{apolo2009}, the upper critical dimension ($d_u=6$) 
\cite{melchert2010}, another upper critical dimension ($d_u^{\rm DPL}=3$) for densely packed 
loops far above the critical point \cite{melchert2011}, the mean-field behavior on a random graph 
with fixed connectivity \cite{mezard2011}, the Schramm-Loewner evolution properties of paths in 2D 
lattices \cite{norrenbrock2013}, and loop-length distributions in several dimensions 
\cite{claussen2012}.

Nevertheless, all this work was for non-directed lattices or graphs. Thus, as compared to the change 
which occurs when moving from standard to directed percolation, it is valid to ask whether the 
directed variant of NWP, which is introduced and studied in this work, gives again rise to a new 
type of behavior.

Note that, while DP is defined as a local growth process, the path-like clusters in NWP emerge due 
to global optimization. This is also true for directed polymers in random media (DPRM) 
\cite{kardar1987}, but unlike NWP, DPRM does not feature a phase transition. Nevertheless, it will 
be outlined in this article that NWP and DPRM are partially related to each other.

Next, we outline the model and the numerical procedures. NWP can be defined on any graph, in 
particular finite-dimensional lattices, which we consider here. In particular we study here a 
directed, weighted, periodic, simple square lattices with side length $L$. The direction of the 
edges is arranged as follows: All horizontal edges point to the left and all vertical ones point up.
Edge weights are drawn from a distribution which provides both positive and negative values. The 
proportion of negative and positive weights can be tuned by a disorder parameter $\rho$.

We study paths and loops in the lattice. For each path or loops a weight is defined which consist 
of the sum of the weights of the edges contributing to the loop or path. For a given realization of 
the disorder, an \emph{optimal} configuration consisting of one path and zero or possibly a finite 
number of system-spanning loops is determined. The configuration must fulfill some constrains: The 
path must be fixed at the bottom right, the loops and the path are not allowed to intersect one 
another, and the total sum of all weights of the loops and the path must be an exact \emph{minimum}.
The algorithm is very similar to the undirected case, the changes which are necessary are explained 
in Sec.\ \ref{sect:modelalgo}.

\begin{figure}[bt]
    \centerline{
        \includegraphics[width=1.\linewidth]{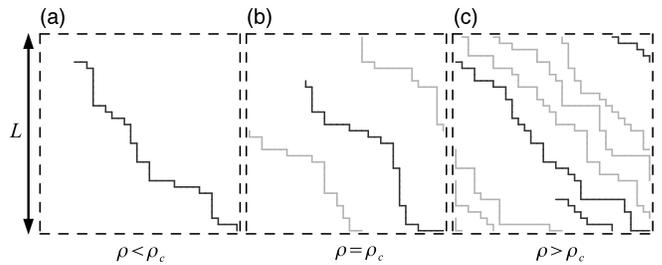}}
        \caption{
        Illustration of minimum-weight configurations consisting of loops (gray) and one path (black) 
        in a directed 2D square lattice of side length $L=32$ with periodic boundary conditions.
        The path is forced to start at the right bottom corner. For small values of $\rho$, there 
        does not appear a loop and also the path does not span the lattice. At $\rho = \rho_c$ one 
        percolating loop occurs. For large values of $\rho$, there are many spanning loops and also 
        the path is percolating. 
        \label{fig:comparison}}
\end{figure}

Fig.\ \ref{fig:comparison} shows such optimal configurations for different values of $\rho$ in a 
lattice of size $L=32$. The main question we are interested in is, whether there is a system 
spanning path or system spanning loops, i.e., whether at least one object percolates. Note that for 
analyzing geometrical properties we are using the path. The loops arise (unavoidably) from the 
global nature of the underlying optimization problem, see the technical details in Sec.\ 
\ref{sect:results} and allow us to study the percolation transition at $\rho=\rho_c$, i.e., whether 
the full lattice admits one or several percolating objects. Note that here, due to the construction 
of the underlying lattice with directed edges, small loops can not occur, in contrast to the 
undirected NWP. In particular, this is true for $\rho<\rho_c$ (cf.\ Fig.\ 
\ref{fig:comparison}(a)) where just a finite path appears and loops are absent. At the critical 
point $\rho=\rho_c$ there appears a lattice-spanning loop in the depicted example Fig.\ 
\ref{fig:comparison}(b), since there are enough negatively weighted edges in the lattice. Also a 
percolating path and no percolating loop might appear for some realizations. Above the critical 
point $\rho>\rho_c$ the number of lattice-spanning loops increases and even the path winds 
around the lattice (cf.\ Fig.\ \ref{fig:comparison}(c)). Thus, this article we study the 
disorder-driven, geometric phase transition and determine its characterizing critical exponents, in 
particular with respect to the parallel and perpendicular correlation lengths.

As an interpretation of the NWP problem, one can imagine an agent that takes a trip in a graph 
along the path.Whenever he travels along a positively weighted edge, the agent has to pay some 
resource according to the positive value. On the other hand, he will harvest some resource, if he 
travels along a negatively weighted edge. Therefore, the optimal path/loop configuration obtained in
the context of the NWP problem provides the optimal route of the agent (path), possibly in 
competition with other agents (loops), to gain as many resources as possible. Only paths or loops 
which lead to a larger amount of harvested resources as compared to the paid resources will occur.

The remainder of this article is organized as follows. In Sec.\ \ref{sect:modelalgo}, we introduce 
the model in more detail and explain the algorithm. In Sec.\ \ref{sect:results}, we describe the 
finite-size scaling technique that has been used to estimate the critical exponents numerically and 
present our results. We close with a summary in Sec. \ref{sect:summary}.

\section{Model and algorithm \label{sect:modelalgo}}
The underlying graph $G=(V,E)$ at hand is a 2D directed square lattice whose edges point either to 
the left or up. Its boundaries are periodic meaning the lattice can be considered as placed on a 
torus in a topological sense. Each edge $e_{ij}\in E$ carries a weight $\omega_{ij}$ that is taken 
from a ``Gauss-like'' distribution characterized  by a tunable disorder parameter $\rho$:
\begin{align}
    P(\omega) = (1-\rho)\delta(\omega-1)+\rho\exp(-&\omega^2)/\sqrt{2\pi},\nonumber\\
    &0\leq\rho\leq 1.
    \label{eq:gausslike}
\end{align}
The shape of the lattice is quadratic in all simulations, so the number of nodes is 
$N=\vert V\vert=L^2$.

Given such a graph and a realization of the disorder, an optimal configuration consisting of an 
arbitrary number of loops, i.e., closed paths, and one additional path (possibly with zero length 
is computed. The configuration must fulfill following requirements: i) One endpoint of the path 
must be pinned at the bottom right corner. However, it is also allowed that no path occurs. ii) The 
loops and the paths are not allowed to cross or touch each other. iii) The configurational energy
\begin{align}
    \mathcal{E} = \sum_{\mathcal{L}\in\mathcal{C}}\omega_{\mathcal{L}}
    \label{eq:energy}
\end{align}
has to be minimized. Here $\omega_\mathcal{L}$ denotes the total of all edge weights belonging to 
loop or path $\mathcal{L}$. Note that Eq.\ \ref{eq:gausslike} provides real numbers, so the optimal 
configuration is unique for each realization of the disorder. Since the number of loops is not 
specified and even the path might not appear (zero length), also an empty configuration might be 
valid. This would be the case, e.g., if all edges carried a positive weight. Loops and also the path 
can solely appear, if their weight is negative, otherwise $\mathcal{E}$ would not be minimal, since 
an empty configuration has $\mathcal{E}=0$. Furthermore, since all edges point either to the 
left or up, loops can appear only, if they span the lattice in either horizontal or vertical 
direction. Therefore, their smallest length is $L$. As a matter of fact, the typical length of the 
loops is $2L$ in the vicinity of the critical point.

\begin{figure}[bt]
    \centerline{
        \includegraphics[width=1.\linewidth]{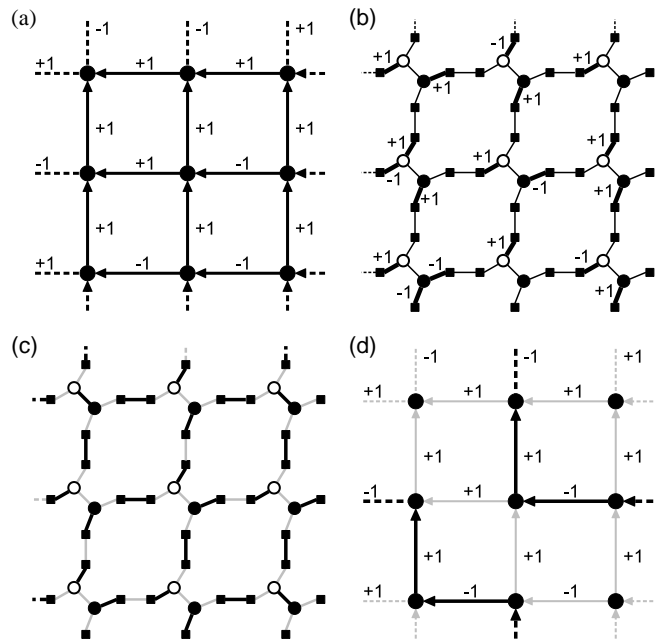}}
        \caption{
        Illustration of the algorithmic procedure for a periodic lattice of size $L=3$. For the 
        sake of clarity, the procedure is just depicted for a directed lattice that does not contain 
        a finite path. In Sec. \ref{sect:modelalgo} it is described how the construction of the 
        auxiliary graph must be altered in order to force a path in the lattice that starts at the 
        bottom right corner and terminates at any node. (a) Original lattice with weighted, directed 
        edges. (b) Auxiliary graph with proper weight assignment. The thick edges carry the weight 
        as the respective edges in the original graph. The weights of all other edges are zero. (c) 
        Illustration of the MWPM: black edges are matched and gray ones are unmatched. For the sake 
        of clarity, edge weights are not depicted. (d) Reconstruction to the original lattice taking 
        the MWPM result into account.
        \label{fig:algorithm}}
\end{figure}

In order to find the optimal configuration, we transform the original graph to an appropriate 
auxiliary graph first. Subsequently, a \emph{minimum-weight perfect matching} (MWPM) 
\cite{melchert2009thesis,cook1999,optphys2001} provides all information to reconstruct the original 
graph exhibiting the correct loop/path configuration. Fig. \ref{fig:algorithm} is an illustration of 
the algorithmic procedure for a given realization of the disorder for a periodic lattice of size 
$L=3$. For reasons of clarity, only loops but no path can occur here.
Guided by Fig. \ref{fig:algorithm}, we give a concise description how the algorithm works. 
Afterwards, we explain in which way the algorithm must be altered, so that the appearance of a path 
becomes possible.

(1) First all original nodes are \emph{duplicated}. For each pair of duplicated nodes, one 
additional edge with zero weight is added linking both nodes of a pair. The two duplicated nodes of 
one pair (represented black and white in Fig. \ref{fig:algorithm}(b)) are treated differently.
Considering just one pair, if an adjacent edge has pointed to the original node, this edge will be 
linked to one of the duplicated nodes (in Fig.\ \ref{fig:algorithm}(b) this node is the black one).
On the other hand, if the edge has pointed away from the original node, it will be connected to the 
other duplicated node (in Fig.\ \ref{fig:algorithm}, this node is white). Subsequently, each of the 
original edges is replaced by a path of three edges and two \emph{additional} nodes (in Fig.\ 
\ref{fig:algorithm}(b), these nodes are depicted as squares). One of the two edges that are 
connected with the duplicates of the original nodes is assigned with the weight of the original 
edge. The other two edges carry zero weight. The resulting auxiliary graph is illustrated in Fig.\ 
\ref{fig:algorithm}(b), where bold edges carry the original weights and the thin ones carry a 
weight of zero. A more extensive and pedagogical description of the mapping (for the undirected 
variant of the model, where the auxiliary graph is slightly different) can be found in Ref.\ 
\cite{melchert2007}.

(2) A MWPM is determined on the auxiliary graph via exact combinatorial optimization algorithms 
\cite{comment_cookrohe}. A perfect matching is a subset of edges $M$ which ensures that each node in 
the graph has exactly one incident edge $\in M$. There are several subsets that fulfill this 
condition. The MWPM is that perfect matching which has the lowest total weight. For the given 
example, the MWPM is illustrated in Fig.\ \ref{fig:algorithm}(c). Edges that belong to the MWPM are 
represented bold and black.

(3) After determining the MWPM, the original graph can be reconstructed. If and only if the edge 
that links two additional nodes (in Fig.\ \ref{fig:algorithm}(c) these nodes are illustrated as 
squares) does belong to the MWPM, the corresponding edge in the original graph is not part of the 
optimal loop/path configuration. If, on the other hand, the two additional nodes are not matched to 
each other, by the definition of the MWPM, they have to be matched to duplicated nodes, 
respectively. In this case the corresponding edge of the original graph is part of a loop. In this 
way the complete optimal loop/path configuration can be determined. In the presented example (cf.\ 
Fig.\ \ref{fig:algorithm}(d)) the optimal configuration consists of one loop with total weight $-2$.

As the algorithm has been presented, it is not possible to find a path that is pinned in the bottom 
right corner. In order to enable such a path, the auxiliary graph must be expanded. After 
constructing the auxiliary graph as described above, the white duplicate of the original node in the 
bottom right corner gets connected via a path consisting of three edges (all carry zero weight) and 
two nodes to the black duplicates of all other original nodes. This means technically the path is 
also a loop, but the ``returning'' part of the loop is ``hidden'' with respect to the original 
lattice, such that it appears as a path there. Such an auxiliary graph is not planar and contains 
many additional edges, therefore, we do not depict this additional specification in the illustration 
Fig.\ \ref{fig:algorithm}.

\section{Results \label{sect:results}}
The NWP model exhibits a geometrical continuous phase transition. For a small amount of negative 
weights, the path would appear rather short and loops would not appear at all, if the system size 
were chosen sufficiently large. This can be seen in Fig.\ \ref{fig:comparison}. For small values of 
$\rho$, the formation of loops is suppressed, because each possible loop has length $O(L)$ and thus
would collect too many positively weighted edges. This is clearly different in the undirected 
variant of the model, where also small loops will appear, even if $\rho$ is small 
\cite{melchert2008}. On the other hand, if $\rho$ is large, the path might grow very long and even
multiple loops will occur.

The two regions, in which lattice-spanning, i.e., percolating, loops or paths will or, respectively, 
will not occur with high probability, are separated by a certain value of $\rho=\rho_c(L)$, the 
critical point. In the thermodynamic limit, i.e., $L\rightarrow\infty$, there are no 
lattice-spanning objects in the lattice, if $\rho<\rho_c=\rho_c(\infty)$. On the other hand, if 
$\rho> \rho_c$, there will appear some percolating objects always.

In this section we determine the critical point and estimate the critical exponents that 
characterize the phase transition via a finite-size scaling analysis. Note that a common scaling 
assumption \cite{stauffer1992} that is typically used for undirected models cannot be applied here.
Therefore, due to the anisotropic nature of the underlying lattice (parallel and perpendicular to 
the natural diagonal orientation), there are two different correlation lengths that have a different 
asymptotic behavior
\begin{align}
    \xi_{\parallel} \sim \vert\rho-\rho_c\vert^{-\nu_\parallel},\hspace{1cm}\xi_{\perp} \sim \vert\rho-\rho_c\vert^{-\nu_\perp}
\end{align}
in the thermodynamic limit, with  $\nu_\parallel$ and $\nu_\perp$ being the critical exponents 
describing the power-law divergence of the correlation lengths, respectively. At the critical point, 
their finite-size scaling is assumed to be \cite{sinha2008}
\begin{eqnarray}
\xi_\parallel & \sim & L^{\theta_\parallel} \nonumber\\
\xi_\perp & \sim & L^{\theta_\perp}\,.
\label{eq:scale:crit}
\end{eqnarray}

For anisotropic percolation models a phenomenological finite-size scaling theory is introduced in 
Ref.\ \cite{sinha2008}. It is expected that cluster related quantities $y(L,\rho)$ can be rescaled 
according to
\begin{align}
    y(L,\rho) &= L^{-b\,\theta_{\parallel}/\nu_{\parallel}}f[(\rho-\rho_c)L^{\theta_{\parallel}/\nu_\parallel}]\nonumber\\
              &= L^{-b\,\theta_{\perp}/\nu_{\perp}}f[(\rho-\rho_c)L^{\theta_{\perp}/\nu_\perp}],
    \label{eq:scalasump}
\end{align}
where $f[\cdot]$ is an unknown scaling function and $b$ represents a dimensionless critical exponent 
that describes the asymptotic behavior of $y(L,\rho)$ in the thermodynamic limit. According to Eq.\ 
\ref{eq:scalasump}, if $\rho_c$, $\theta_\parallel/\nu_\parallel$ and $b$ are chosen properly, all 
data points of $y(L,\rho)L^{b\,\theta_{\parallel}/\nu_{\parallel}}$ have to lie on one single curve.
Therefore, $y(L,\rho)$ can be measured numerically for different values of $L$ and $\rho$ and, 
subsequently, $y(L,\rho)L^{b\,\theta_{\parallel}/\nu_{\parallel}}$ can be plotted against 
$(\rho-\rho_c)L^{\theta_{\parallel}/\nu_\parallel}$. Then, the unknown constants $\rho_c$, 
$\theta_\parallel/\nu_\parallel$ and $b$ can be adjusted until the data ``collapses'' to one curve 
indicating that the correct values of the constants are found. The same also applies for 
$\theta_\perp/\nu_\perp$ instead of $\theta_\parallel/\nu_\parallel$. Note, that Eq.\ 
\ref{eq:scalasump} shows the scaling behavior of systems that are sufficiently large only 
\cite{binder2002}. All data collapses in this article are made with a computer-assisted scaling 
analysis \cite{autoScale2009}.

\begin{figure}[bt]
    \centerline{
        \includegraphics[width=1.\linewidth]{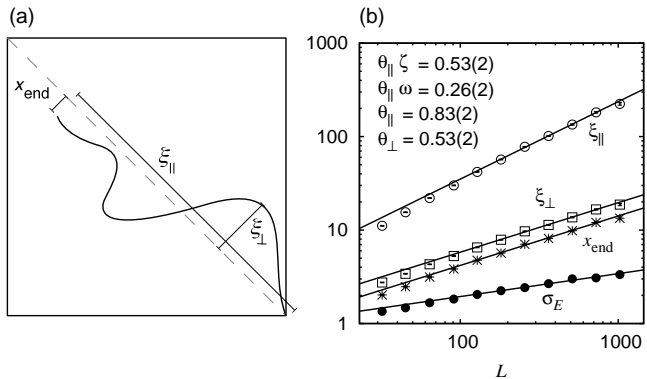}}
        \caption{
        (a) Sketch of $\xi_\parallel$, $\xi_\perp$ and $x_{\rm end}$. (b) Plot shows $\xi_\parallel$ 
        (red.\ $\chi^2=1.1$), $\xi_\perp$ (red.\ $\chi^2=1.5$), $x_{\rm end}$ (red.\ 
        $\chi^2=2.1$) and $\sigma_E$ (red.\ $\chi^2=3.8$) as a function of $L$. Merely 
        system sizes from $L=181$ to $724$ have been considered for the power-law regression 
        curves. The measurements are taken at the estimated value of the critical point 
        $\rho_c=0.3789$.
        \label{fig:theta}}
\end{figure}

This data collapse approach allows only to determine the ratios $\theta_\perp/\nu_\perp$ and 
$\theta_\parallel/\nu_\parallel$. In order to find an estimate for $\nu_\parallel$ and $\nu_\perp$,
we additionally determine $\theta_\parallel$ and $\theta_\perp$ directly by applying Eqs.\ 
\ref{eq:scale:crit}. For that reason, the path is forced on the lattice, because, as evident from 
Fig.\ \ref{fig:theta}(a), the correlation lengths can be estimated by taking measurements of the 
path. The measurements are taken at the estimated critical point $\rho_c=0.3789$ that has been 
found with the data collapse technique described above and will be presented below. For the ease of 
presentation, we do not have to deal with the ratios $\theta/\nu$, we have switched the order here 
and show the determination of $\theta_\parallel$  and $\theta_\perp$ first. Fig.\ \ref{fig:theta}(b) 
shows that a very clean power law behavior is visible, leading to $\theta_\parallel =0.83(2)$ 
and $\theta_\perp=0.53(2)$. 

To actually determine the critical point and obtain the other critical exponents, we have monitored 
several observables in the vicinity of the expected value of the critical point 
$(p\in[0.377,0.382])$ for different system sizes. Since we could use fast optimization algorithms, 
we could study rather large system sizes in the range $L=256$ to $L=724$ with good statistics: The 
data have been obtained by averaging over $20000$ ($L=256$), $16000$ ($L=362$), $10000$ 
($L=512$) and $8000$ ($L=724$) realizations of the disorder, respectively.
\begin{figure}[bt]
    \centerline{
        \includegraphics[width=1.\linewidth]{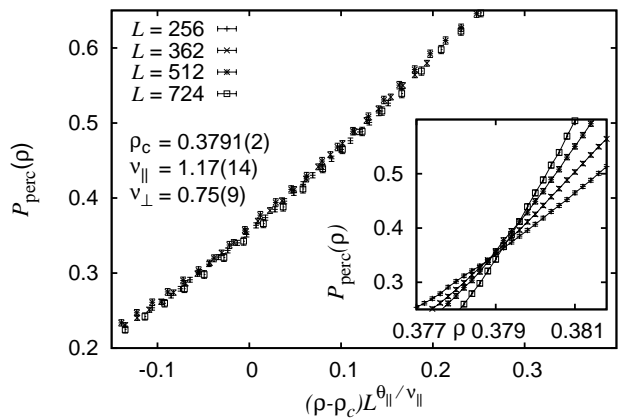}}
        \caption{
        Percolation probability $P_{\rm perc}(\rho)$ as a function of $\rho$ in the vicinity of the 
        critical point (inset). The data is collapsed to one curve by using the scaling assumption 
        Eq.\ \ref{eq:scalasump} (main plot).       
        \label{fig:percProb}}
\end{figure}

Fig.\ \ref{fig:percProb} shows the percolation probability $P_{\rm perc}(\rho)$ as a function of the 
disorder parameter $\rho$ as well as the rescaled data collapse. Since the percolation probability 
is a dimensionless quantity, $b=0$ is set in Eq.\ \ref{eq:scalasump}. The estimates 
$\rho_c=0.3791(2)$, $\nu_\parallel=1.17(14)$ and $\nu_\perp=0.75(9)$ provide the best 
data collapse with quality $S=1.2$, which denotes the mean-square distance of the data points to 
the unknown scaling function in units of the standard error \cite{autoScale2009}.
\begin{figure}[bt]
    \centerline{
        \includegraphics[width=1.\linewidth]{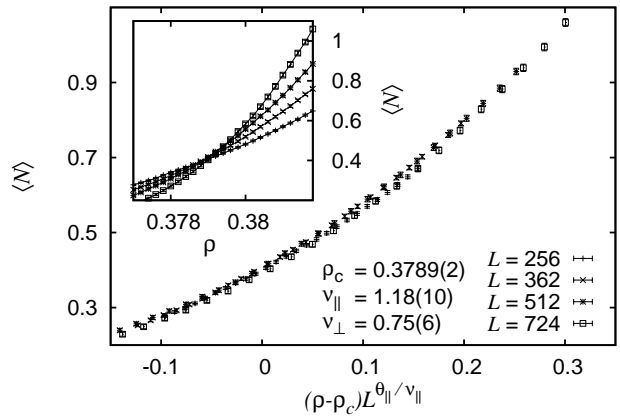}}
        \caption{
        Average number of spanning loops $\langle N\rangle$ as a function of $\rho$ in the vicinity 
        of the critical point (inset). The data is collapsed to one curve by using the scaling 
        assumption Eq.\ \ref{eq:scalasump} (main plot).  
        \label{fig:numloops}}
\end{figure}

We have also measured the average number of lattice-spanning objects $\langle N\rangle$, see Fig.\ 
\ref{fig:numloops}. Note that, in contrast to standard percolation, more than one object can be 
spanning. By using again the data-collapse approach, we have found $\rho_c=0.3789(2)$, 
$\nu_\parallel=1.18(10)$ and $\nu_\perp=0.75(6)$ with quality $S=2.3$.

\begin{figure}[bt]
    \centerline{
        \includegraphics[width=1.\linewidth]{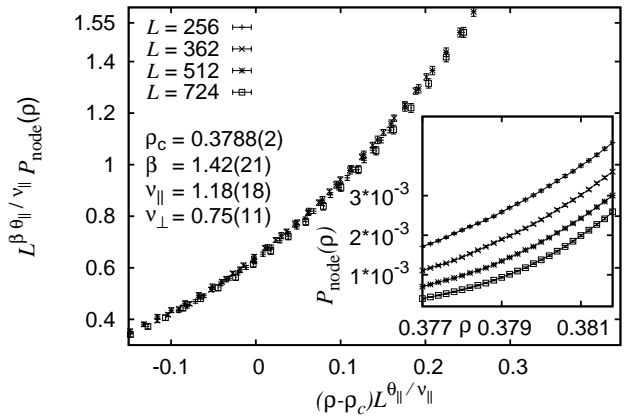}}
        \caption{
        Order parameter $P_{\rm node}(\rho)$ as a function of $\rho$ in the vicinity of the 
        critical point (inset). The data is collapsed to one curve by using the scaling assumption 
        Eq.\ \ref{eq:scalasump} (main plot).
        \label{fig:orderpara}}
\end{figure}

Another quantity that has been under scrutiny is the order parameter
\begin{align}
    P_{\rm node} \equiv \frac{\langle l\rangle}{L^d},
    \label{eq:orderpara}
\end{align}
which is the probability that an edge belongs to either a percolating loop or percolating path. The 
total number of all edges that belong to the percolating objects is given by $l$. $d=2$ 
signifies the dimension of the lattice. The asymptotic behavior of the order parameter is governed 
by an additional critical exponent $\beta$, the percolation strength \cite{stauffer1994}. As evident 
from Fig.\ \ref{fig:orderpara}, we have found $\rho_c=0.3788(2)$, $\nu_\parallel=1.18(18)$, 
$\nu_\perp=0.75(11)$ and $\beta=1.42(21)$ with quality $S=1.3$.
It should be noted that several combinations of $\rho_c$ and the 
exponents provide valid data collapses. Therefore,
we considered $P_{\rm node}$ 
versus $L$ at the critical point (plot not shown here), which exhibits only
one fitting parameter. We found for large
system sizes a power law behavior, which is compatible with
$\beta=1.42(21)$, which we therefore take as final estimate. 

\begin{figure}[bt]
    \centerline{
        \includegraphics[width=1.\linewidth]{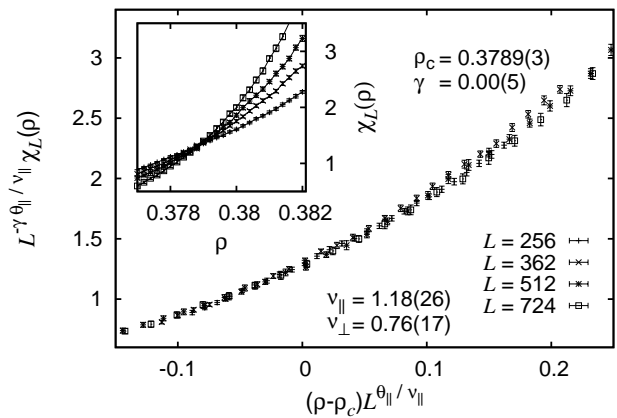}}
        \caption{Fluctuations of the order parameter $\chi_L(\rho)$ over $\rho$ in the vicinity of 
        the critical point (inset). The data is collapsed to one curve by using the scaling 
        assumption Eq.\ \ref{eq:scalasump} (main plot).
        \label{fig:suscept}}
\end{figure}

Next, we consider the associated finite-size susceptibility
\begin{align}
    \chi_L = L^{-d}(\langle l^2\rangle-\langle l\rangle^2),
    \label{eq:suscept}
\end{align}
whose asymptotic behavior is guided by the critical exponent $\gamma$. As can be seen from Fig.\ 
\ref{fig:suscept}, the best data collapse is provided by $\rho_c=0.3789(3)$, 
$\nu_\parallel=1.18(26)$, $\nu_\perp=0.76(17)$ and $\gamma=0.00(5)$ with quality 
$S=0.8$.

\begin{figure}[bt]
    \centerline{
        \includegraphics[width=1.\linewidth]{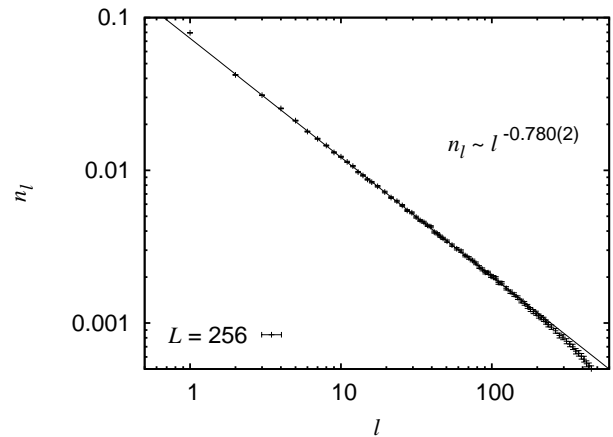}}
        \caption{
        Distribution of the path lengths $l$ at the critical point excluding those which percolate. 
        $1200000$ realizations of the disorder have been considered. For the fit (red.\ 
        $\chi^2=1.0$) path lengths from $l=2$ to $100$ have been taken into account only. 
        \label{fig:clusterDist}}
\end{figure}

Right at the critical point, we studied the distribution of path-lengths excluding the 
lattice-spanning ones. As evident from Fig.\ \ref{fig:clusterDist}, the distribution is in good 
agreement with a power law decay $n_l\sim l^{-\tau}$ with $\tau=0.780(2)$.

NWP is defined as a global optimization problem (cf.\ Sec.\ \ref{sect:modelalgo}) to find the 
minimally weighted configuration consisting of loops plus one path. Since in the directed polymer 
problem \cite{kardar1987} minimally weighted paths are also selected by global optimization in a 
random media, these two models might be related. The DPRM can be described as follows: A weighted 
square lattice, in which all edges carry a positive weight, gets cut along its diagonal and then it 
is oriented as a triangle, whose right angle is up. Then all edges become directed and point either 
to bottom right or bottom left. On such a lattice, for a given realization of the disorder, one 
looks for the minimally weighted path that goes from the apex to the base. It has been shown in 
Ref.\ \cite{huse1985} that $\zeta^\circ=2/3$, where $\zeta^\circ$ is the roughness exponent 
defined by $D\sim t^{\zeta^\circ}$. $D$ describes the mean distance between the base center and the 
endpoint of the path and $t$ is the size of the triangle. Furthermore, it is also shown in Ref.\ 
\cite{huse1985} that $\omega^\circ=1/3$, which is defined by $\sigma_E\sim t^{\omega^\circ}$. 
$\sigma_E$ denotes the standard deviation of the weight of the optimal path. A relation between 
these two exponents is given by the scaling relation $\omega^\circ=2\zeta^\circ-1$ 
\cite{huse1985}. There are some differences between the optimal path in DPRM and the path that 
appears in directed NWP. First of all, the directed NWP includes the disorder parameter $\rho$ 
which allows us to investigate a percolation transition, which is completely absent for DPRM, since 
all paths are system spanning by construction. There are also smaller technical differences: For the 
directed NWP, one looks for an optimal configuration of loops plus one path in NWP. The loops, which 
cannot be crossed by the path, have to be negatively weighted as well and, therefore, block several 
negatively weighted edges that cannot be picked up by the path. Thus, this path can not be 
considered as optimal on its own. Furthermore, while the lengths of the paths in DPRM are always 
equal, the lengths of the paths in NWP differ considerably.

Nevertheless, in spite of one big and the two smaller differences, directed NWP and DPRM exhibit 
some scaling which is comparable. This is not unnatural, since both models describe some optimal
paths in disordered lattices: In order to compare both models, we identify 
$D\leftrightarrow x_{\rm end}$, where $x_{\rm end}$ is the distance between the endpoint of the 
path and the line of predominant direction (cf.\ Fig.\ \ref{fig:theta}(a)) and 
$t\leftrightarrow \xi_\parallel$. Then we consider 
$x_{\rm end}\sim\xi_\parallel^{\,\zeta}\sim L^{\theta_\parallel\zeta}$ and 
$\sigma_E\sim\xi_\parallel^{\,\omega}\sim L^{\theta_\parallel\omega}$ 
for the NWP model. As evident from Fig.\ \ref{fig:theta}(b), $x_{\rm end}$ scales with 
$\theta_\parallel\zeta=0.53(2)$ and $\sigma_E$ with $\theta_\parallel\omega=0.26(2)$.
Consequently, $\omega=0.31(3)$ and $\zeta=0.64(4)$, which are in good agreement with the 
exponents of the DPRM model.

Note that in our model the path is included to determine its geometrical properties, in particular 
its extension parallel and perpendicular to the preferred (diagonal) lattice direction. 
Nevertheless, we have also performed simulations for the model \emph{without} a path, just to study 
the percolation properties of loops alone. All results (for somehow smaller system sizes, not shown 
here) for the percolation properties remain the same within error bars.

\section{Summary \label{sect:summary}} 
In this work with have studied
the directed variant of the  negative-weight percolation model. This
model defined as a global optimization problem. The model can be
studied numerically efficiently,  since a mapping to the
minimum-weight perfect matching problem exist, such that fast
polynomial-time  optimization algorithms can be applied. Thus, large
systems can be studied numerically with good  statistics giving rise
to high-quality results. The model exhibits a continuous phase
transition,  that is characterized by the appearance of loops and a
path where at least one of them is large,  i.e., system-spanning. We
have studied this percolation transition by extended numerical
simulations  and their analysis based on a finite-size scaling
method. By investigating several cluster-related  observables we found
estimates for the percolation threshold, which we summarized here as
$\rho_c=0.3789(3)$, several critical exponents
$\nu_\parallel=1.18(10)$,  $\nu_\perp=0.75(6)$,
$\beta=1.42(21)$, $\gamma=0.00(5)$ and an exponent that
describes  the power-law decay of the path-length distribution
$\tau=0.780(2)$. For the values of the correlation lengths,
we have taken the estimates which yielded the smallest statistical error
bars (from the data collapse of the average number $\langle N\rangle$
of percolating loops). These values are compatible with the estimates 
from the scaling of other quantities. Finally, we tested the scaling relation 
  $2\beta=\nu_\parallel+\nu_\perp-\gamma$
\cite{essam1988}, which is a standard relation for directed percolation.
For the left side we get $2.84(42)$ while for the right side we get
$1.93(21)$. Thus within one-sigma, the scaling relation is not fulfilled,
while within two-sigma, the left and right side are compatible. Thus, 
it is presently not fully clear whether the scaling relation is fulfilled.
If not, it could be due to the fact that the percolating objects are
line-like rather than bulk-like. Note  that for standard directed percolation
near a wall, for the results obtained using a series expansion the
 scaling relation is clearly violated \cite{essam1996}.
Nevertheless, in the case of a violation it would be 
different from the undirected NWP case, where the standard scaling relations
for percolation hold \cite{melchert2008,melchert2010}.

Additionally, we have shown that the directed negative-weight percolation model is related to 
directed polymers in random media (DPRM), although the DPRM does not exhibit a percolation
transition (except when diluting the system where just the standard percolation transition appears.)

\acknowledgments
    Financial support was obtained via the Lower Saxony research network ``Smart Nord'' which 
    acknowledges the support of the Lower Saxony Ministry of Science and Culture through the 
    ``Nieders\"achsisches Vorab'' grant program (grant ZN 2764/ ZN 2896). The simulations  were 
    performed at the \emph{HERO} cluster for scientific computing of the University of Oldenburg 
    jointly funded by the DFG (INST 184/108-1 FUGG) and the ministry of Science and Culture (MWK) 
    of the Lower Saxony State.

\bibliography{research.bib}

\begin{thebibliography}{26}
\expandafter\ifx\csname natexlab\endcsname\relax\def\natexlab#1{#1}\fi
\expandafter\ifx\csname bibnamefont\endcsname\relax
  \def\bibnamefont#1{#1}\fi
\expandafter\ifx\csname bibfnamefont\endcsname\relax
  \def\bibfnamefont#1{#1}\fi
\expandafter\ifx\csname citenamefont\endcsname\relax
  \def\citenamefont#1{#1}\fi
\expandafter\ifx\csname url\endcsname\relax
  \def\url#1{\texttt{#1}}\fi
\expandafter\ifx\csname urlprefix\endcsname\relax\def\urlprefix{URL }\fi
\providecommand{\bibinfo}[2]{#2}
\providecommand{\eprint}[2][]{\url{#2}}

\bibitem[{\citenamefont{Stanley}(1999)}]{stanley1999}
\bibinfo{author}{\bibfnamefont{H.~E.} \bibnamefont{Stanley}},
  \bibinfo{journal}{Rev. Mod. Phys.} \textbf{\bibinfo{volume}{71}},
  \bibinfo{pages}{358} (\bibinfo{year}{1999}).

\bibitem[{\citenamefont{Stauffer}(1979)}]{stauffer1979}
\bibinfo{author}{\bibfnamefont{D.}~\bibnamefont{Stauffer}},
  \bibinfo{journal}{Physics Reports} \textbf{\bibinfo{volume}{54}},
  \bibinfo{pages}{1} (\bibinfo{year}{1979}).

\bibitem[{\citenamefont{Stauffer and Aharony}(1992)}]{stauffer1992}
\bibinfo{author}{\bibfnamefont{D.}~\bibnamefont{Stauffer}} \bibnamefont{and}
  \bibinfo{author}{\bibfnamefont{A.}~\bibnamefont{Aharony}},
  \emph{\bibinfo{title}{{Introduction to Percolation Theory}}}
  (\bibinfo{publisher}{Taylor \& Francis}, \bibinfo{year}{1992}).

\bibitem[{\citenamefont{Fortuin and Kasteleyn}(1972)}]{fortuin1972}
\bibinfo{author}{\bibfnamefont{C.~M.} \bibnamefont{Fortuin}} \bibnamefont{and}
  \bibinfo{author}{\bibfnamefont{P.~W.} \bibnamefont{Kasteleyn}},
  \bibinfo{journal}{Physica} \textbf{\bibinfo{volume}{57}},
  \bibinfo{pages}{536} (\bibinfo{year}{1972}).

\bibitem[{\citenamefont{Hinrichsen}(2000)}]{hinrichsen2000}
\bibinfo{author}{\bibfnamefont{H.}~\bibnamefont{Hinrichsen}},
  \bibinfo{journal}{Adv. Phys.} \textbf{\bibinfo{volume}{49}},
  \bibinfo{pages}{815} (\bibinfo{year}{2000}).

\bibitem[{\citenamefont{Melchert and Hartmann}(2008)}]{melchert2008}
\bibinfo{author}{\bibfnamefont{O.}~\bibnamefont{Melchert}} \bibnamefont{and}
  \bibinfo{author}{\bibfnamefont{A.~K.} \bibnamefont{Hartmann}},
  \bibinfo{journal}{New. J. Phys.} \textbf{\bibinfo{volume}{10}},
  \bibinfo{pages}{043039} (\bibinfo{year}{2008}).

\bibitem[{\citenamefont{Apolo et~al.}(2009)\citenamefont{Apolo, Melchert, and
  Hartmann}}]{apolo2009}
\bibinfo{author}{\bibfnamefont{L.}~\bibnamefont{Apolo}},
  \bibinfo{author}{\bibfnamefont{O.}~\bibnamefont{Melchert}}, \bibnamefont{and}
  \bibinfo{author}{\bibfnamefont{A.~K.} \bibnamefont{Hartmann}},
  \bibinfo{journal}{Phys. Rev. E} \textbf{\bibinfo{volume}{79}},
  \bibinfo{pages}{031103} (\bibinfo{year}{2009}).

\bibitem[{\citenamefont{Melchert et~al.}(2010)\citenamefont{Melchert, Apolo,
  and Hartmann}}]{melchert2010}
\bibinfo{author}{\bibfnamefont{O.}~\bibnamefont{Melchert}},
  \bibinfo{author}{\bibfnamefont{L.}~\bibnamefont{Apolo}}, \bibnamefont{and}
  \bibinfo{author}{\bibfnamefont{A.~K.} \bibnamefont{Hartmann}},
  \bibinfo{journal}{Phys. Rev. E} \textbf{\bibinfo{volume}{81}},
  \bibinfo{pages}{051108} (\bibinfo{year}{2010}).

\bibitem[{\citenamefont{Melchert and Hartmann}(2011)}]{melchert2011}
\bibinfo{author}{\bibfnamefont{O.}~\bibnamefont{Melchert}} \bibnamefont{and}
  \bibinfo{author}{\bibfnamefont{A.~K.} \bibnamefont{Hartmann}},
  \bibinfo{journal}{Eur. Phys. J. B} \textbf{\bibinfo{volume}{80}},
  \bibinfo{pages}{155} (\bibinfo{year}{2011}).

\bibitem[{\citenamefont{Melchert et~al.}(2011)\citenamefont{Melchert, Hartmann,
  and M\'ezard}}]{mezard2011}
\bibinfo{author}{\bibfnamefont{O.}~\bibnamefont{Melchert}},
  \bibinfo{author}{\bibfnamefont{A.~K.} \bibnamefont{Hartmann}},
  \bibnamefont{and} \bibinfo{author}{\bibfnamefont{M.}~\bibnamefont{M\'ezard}},
  \bibinfo{journal}{Phys. Rev. E} \textbf{\bibinfo{volume}{84}},
  \bibinfo{pages}{041106} (\bibinfo{year}{2011}).

\bibitem[{\citenamefont{Claussen et~al.}(2012)\citenamefont{Claussen, Melchert,
  and Hartmann}}]{claussen2012}
\bibinfo{author}{\bibfnamefont{G.}~\bibnamefont{Claussen}},
  \bibinfo{author}{\bibfnamefont{O.}~\bibnamefont{Melchert}}, \bibnamefont{and}
  \bibinfo{author}{\bibfnamefont{A.~K.} \bibnamefont{Hartmann}},
  \bibinfo{journal}{Phys. Rev. E} \textbf{\bibinfo{volume}{86}},
  \bibinfo{pages}{056708} (\bibinfo{year}{2012}).

\bibitem[{\citenamefont{Norrenbrock et~al.}(2013)\citenamefont{Norrenbrock,
  Melchert, and Hartmann}}]{norrenbrock2013}
\bibinfo{author}{\bibfnamefont{C.}~\bibnamefont{Norrenbrock}},
  \bibinfo{author}{\bibfnamefont{O.}~\bibnamefont{Melchert}}, \bibnamefont{and}
  \bibinfo{author}{\bibfnamefont{A.~K.} \bibnamefont{Hartmann}},
  \bibinfo{journal}{Phys. Rev. E} \textbf{\bibinfo{volume}{87}},
  \bibinfo{pages}{032142} (\bibinfo{year}{2013}).

\bibitem[{\citenamefont{Melchert et~al.}(2014)\citenamefont{Melchert,
  Norrenbrock, and Hartmann}}]{melchert2014}
\bibinfo{author}{\bibfnamefont{O.}~\bibnamefont{Melchert}},
  \bibinfo{author}{\bibfnamefont{C.}~\bibnamefont{Norrenbrock}},
  \bibnamefont{and} \bibinfo{author}{\bibfnamefont{A.~K.}
  \bibnamefont{Hartmann}}, \bibinfo{journal}{Physics Procedia}
  \textbf{\bibinfo{volume}{57}}, \bibinfo{pages}{58} (\bibinfo{year}{2014}).

\bibitem[{\citenamefont{Kardar and Zhang}(1987)}]{kardar1987}
\bibinfo{author}{\bibfnamefont{M.}~\bibnamefont{Kardar}} \bibnamefont{and}
  \bibinfo{author}{\bibfnamefont{Y.-C.} \bibnamefont{Zhang}},
  \bibinfo{journal}{Phys. Rev. Lett.} \textbf{\bibinfo{volume}{58}},
  \bibinfo{pages}{2087} (\bibinfo{year}{1987}).

\bibitem[{\citenamefont{Melchert}(2009{\natexlab{a}})}]{melchert2009thesis}
\bibinfo{author}{\bibfnamefont{O.}~\bibnamefont{Melchert}},
  \emph{\bibinfo{title}{{PhD thesis}}} (\bibinfo{publisher}{not published},
  \bibinfo{year}{2009}{\natexlab{a}}).

\bibitem[{\citenamefont{Cook and Rohe}(1999)}]{cook1999}
\bibinfo{author}{\bibfnamefont{W.}~\bibnamefont{Cook}} \bibnamefont{and}
  \bibinfo{author}{\bibfnamefont{A.}~\bibnamefont{Rohe}},
  \bibinfo{journal}{INFORMS J. Computing} \textbf{\bibinfo{volume}{11}},
  \bibinfo{pages}{138} (\bibinfo{year}{1999}).

\bibitem[{\citenamefont{Hartmann and Rieger}(2001)}]{optphys2001}
\bibinfo{author}{\bibfnamefont{A.~K.} \bibnamefont{Hartmann}} \bibnamefont{and}
  \bibinfo{author}{\bibfnamefont{H.}~\bibnamefont{Rieger}},
  \emph{\bibinfo{title}{{Optimization Algorithms in Physics}}}
  (\bibinfo{publisher}{Wiley-VCH}, \bibinfo{address}{Weinheim},
  \bibinfo{year}{2001}).

\bibitem[{\citenamefont{Melchert and Hartmann}(2007)}]{melchert2007}
\bibinfo{author}{\bibfnamefont{O.}~\bibnamefont{Melchert}} \bibnamefont{and}
  \bibinfo{author}{\bibfnamefont{A.~K.} \bibnamefont{Hartmann}},
  \bibinfo{journal}{Phys. Rev. B} \textbf{\bibinfo{volume}{76}},
  \bibinfo{pages}{174411} (\bibinfo{year}{2007}).

\bibitem[{com()}]{comment_cookrohe}
\bibinfo{note}{For the calculation of minimum-weighted perfect matchings we use
  Cook and Rohes blossom4 extension to the Concorde library.},
  \urlprefix\url{http://www2.isye.gatech.edu/~wcook/blossom4/}.

\bibitem[{\citenamefont{Sinha and Santra}(2008)}]{sinha2008}
\bibinfo{author}{\bibfnamefont{S.}~\bibnamefont{Sinha}} \bibnamefont{and}
  \bibinfo{author}{\bibfnamefont{S.~B.} \bibnamefont{Santra}},
  \bibinfo{journal}{Preprint: arXiv:0807.2300v1}  (\bibinfo{year}{2008}).

\bibitem[{\citenamefont{Binder and Herrmann}(2002)}]{binder2002}
\bibinfo{author}{\bibfnamefont{K.}~\bibnamefont{Binder}} \bibnamefont{and}
  \bibinfo{author}{\bibfnamefont{D.~W.} \bibnamefont{Herrmann}},
  \emph{\bibinfo{title}{Monte Carlo Simulation in Statistical Physics: An
  Introduction}} (\bibinfo{publisher}{Springer}, \bibinfo{address}{Berlin},
  \bibinfo{year}{2002}).

\bibitem[{\citenamefont{Melchert}(2009{\natexlab{b}})}]{autoScale2009}
\bibinfo{author}{\bibfnamefont{O.}~\bibnamefont{Melchert}},
  \bibinfo{journal}{Preprint: arXiv:0910.5403v1}
  (\bibinfo{year}{2009}{\natexlab{b}}).

\bibitem[{\citenamefont{Stauffer and Aharony}(1994)}]{stauffer1994}
\bibinfo{author}{\bibfnamefont{D.}~\bibnamefont{Stauffer}} \bibnamefont{and}
  \bibinfo{author}{\bibfnamefont{A.}~\bibnamefont{Aharony}},
  \emph{\bibinfo{title}{{Introduction to Percolation Theory}}}
  (\bibinfo{publisher}{Taylor and Francis, London}, \bibinfo{year}{1994}).

\bibitem[{\citenamefont{Huse and Henley}(1985)}]{huse1985}
\bibinfo{author}{\bibfnamefont{D.~A.} \bibnamefont{Huse}} \bibnamefont{and}
  \bibinfo{author}{\bibfnamefont{C.}~\bibnamefont{Henley}},
  \bibinfo{journal}{Phys. Rev. Lett.} \textbf{\bibinfo{volume}{54}},
  \bibinfo{pages}{2708} (\bibinfo{year}{1985}).

\bibitem[{\citenamefont{Essam et~al.}(1988)\citenamefont{Essam, Guttmann, and
  K.}}]{essam1988}
\bibinfo{author}{\bibfnamefont{J.~W.} \bibnamefont{Essam}},
  \bibinfo{author}{\bibfnamefont{A.~J.} \bibnamefont{Guttmann}},
  \bibnamefont{and} \bibinfo{author}{\bibfnamefont{D.}~\bibnamefont{K.}},
  \bibinfo{journal}{J. Phys. A} \textbf{\bibinfo{volume}{21}},
  \bibinfo{pages}{3815} (\bibinfo{year}{1988}).

\bibitem[{\citenamefont{J.~W.~Essam and TanlaKishani}(1996)}]{essam1996}
\bibinfo{author}{\bibfnamefont{I.~J.} \bibnamefont{J.~W.~Essam},
  \bibfnamefont{A.~J.~Guttmann}} \bibnamefont{and}
  \bibinfo{author}{\bibfnamefont{D.}~\bibnamefont{TanlaKishani}},
  \bibinfo{journal}{J. Physics A} \textbf{\bibinfo{volume}{29}},
  \bibinfo{pages}{1619} (\bibinfo{year}{1996}).

\end{thebibliography}

\end{document}